\documentclass[aps,superscriptaddress,nofootinbib,twocolumn]{revtex4}
\usepackage{amsfonts}
\usepackage{amsmath}
\usepackage{amssymb}
\usepackage{graphicx}

\begin{document}

\author{R. Gonz\'{a}lez Felipe}
\email{gonzalez@cftp.ist.utl.pt}
\affiliation{Instituto Superior de Engenharia de Lisboa\\
Rua Conselheiro Em\'{\i}dio Navarro, 1959-007 Lisboa, Portugal}
\affiliation{Centro de F\'{\i}sica Te\'{o}rica de Part\'{\i}culas, Instituto Superior T\'{e}cnico\\
Avenida Rovisco Pais, 1049-001 Lisboa, Portugal}

\author{A. P\'{e}rez Mart\'{\i}nez}
\email{aurora@icmf.inf.cu}
\affiliation{Instituto de Cibern\'{e}tica Matem\'{a}tica y F\'{\i}sica (ICIMAF) \\
Calle E esq 15 No. 309 Vedado, Havana, 10400, Cuba}

\author{H. P\'{e}rez Rojas}
\email{hugo@icmf.inf.cu}
\affiliation{Instituto de Cibern\'{e}tica Matem\'{a}tica y F\'{\i}sica (ICIMAF) \\
Calle E esq 15 No. 309 Vedado, Havana, 10400, Cuba}

\author{M. Orsaria}
\email{orsaria@cbpf.br} \affiliation{Centro Latinoamericano de
F\'{\i}sica (CLAF)\\ Avenida Venceslau Br\'{a}z 71 Fundos, 22290-140,
Rio de Janeiro, Brazil}

\title{Magnetized Strange Quark Matter and Magnetized Strange Quark Stars}

\begin{abstract}
Strange quark matter could be found in the core of neutron stars or forming strange quark stars. As is well known, these astrophysical objects are endowed with strong magnetic fields which affect the microscopic properties of matter and modify the macroscopic properties of  the system. In this paper we study the role of a strong magnetic field in the thermodynamical properties of a magnetized degenerate strange quark gas, taking into account $\beta$-equilibrium and charge  neutrality. Quarks and electrons interact with the magnetic field via their electric charges and anomalous magnetic moments. In contrast to the magnetic field value of $10^{19}\,$G, obtained when anomalous magnetic moments are not taken into account, we find the upper bound $B \lesssim 8.6 \times 10^{17}\,$G, for  the stability of the system. A phase transition could be hidden for fields greater than this value.
\end{abstract}

\maketitle

\section{Introduction}
\label{sec1}

There are two possibilities for the occurrence of a phase transition between hadronic and strange quark matter (SQM) that are well known. The first could occur at very high temperatures and very low baryon density in the early Universe, and the second, as suggested by Bodmer~\cite{Bodmer:1971we}, at densities of higher order than the nuclear density $n_0 \sim 0.16$~fm$^{-3}$. This phase transition would occur in the Universe, every time that a massive star explodes as a supernova, with its consequent remnant. If Bodmer's conjecture is true, SQM could be succeeded in the inner core of neutron stars where strange quarks would be produced through the weak processes with a dynamical chemical equilibrium among the constituents. It is also possible that  after a supernova explosion its core forms directly a strange quark star (SQS) \cite{Hatsuda:1987ck,Sato:1987rd}.

The key property of SQM is that it has a binding energy that could be lower than that of $^{56}$Fe over a rather wide range of the QCD parameters \cite{Witten:1984rs,Farhi:1984qu}. Thus, it is worthwhile to seek connections between SQM or SQS that could explain the observations of anomalous radiations from anomalous X-pulsars (AXPs) and low energy $\gamma$-ray radiation from soft gamma ray repeaters (SGRs) \cite{Lugones:2002va}.  In several SQM studies the essential conclusion is that a more compact matter would be the cause of these observations~\cite{Horvath:2004gn}. On the other hand, astrophysical observations point out that compact objects are endowed with strong magnetic fields which should play an important role in  neutron stars or SQS. It is believed that magnetic fields larger than $10^{14}$~G are the central engine of their radiations.

As is well known, a magnetic field modifies the microscopic properties of matter with the corresponding macroscopic implication. However, the role of the magnetic field in SQM has not been fully studied and understood. Ref.~\cite{Chakrabarty:1996te} is a pioneer work in this field. The thermodynamical properties of quark matter in a strong magnetic field have also been studied in Refs. \cite{Perez Martinez:2005av,Perez Martinez:2007}, using the MIT bag model modified by the inclusion of the magnetic field in the Lagrangian. In these works, it has been confirmed that there is an anisotropy of pressures due to the strong magnetic field~\cite{Chaichian:1999gd,Martinez:2003dz}, and that the MIT bag model can be used satisfactorily to study the magnetized quark matter. A first approach to consider the role of the anomalous magnetic moment (AMM) for quark matter has been followed in Ref.~\cite{Perez Martinez:2007}.

There are theoretical and experimental studies which indicate that quarks have an AMM \cite{Fajfer:1983iu,Singh:1985sg,Brekke:1994ne,Kopp:1994qv,Bicudo:1998qb}. A stringent bound on the quark AMM has been obtained~\cite{Kopp:1994qv} from high precision measurements at LEP, SLC and HERA.  Thus, the contribution from the AMM of the quarks could be significant in SQM. On the other hand, for electrons, the effects of the AMM turn out to be small over the range of magnetic fields typically attained in neutron stars ($10^{15} - 10^{19}$~G) ~\cite{Duncan:2000pj} and, therefore, can be safely neglected in our calculations.

The scope of the present paper is to study the role of the AMM in the spectrum of particles and its relevance  in the thermodynamic properties and stability of the SQM, considering $\beta$-equilibrium and charge neutrality.  We consider that the constituents of SQM interact with the magnetic field via their charges and their AMM. The effects of this interaction with the AMM could be seen in two forms. The first one is that quarks and electrons have  the energy quantized in Landau levels  due to their charge. The normal magnetic moment is included in the spectrum through an integer moment. This quantization leads to a description of the polarization of the particles and to a softer equation of state (EOS). The second one is due to the inclusion of AMM terms. This causes a further splitting of the energy levels and the spectrum of the particles is no longer degenerate. But the most relevant implication of AMM in the spectrum of the particles comes from the ground state: it could be zero depending on the magnetic field strength. Thus, the AMM for individual particles establishes upper bounds on the strength of the magnetic field. We analyze the consequence of this fact in the macroscopic properties of the SQM in $\beta$-equilibrium.

For individual constituents, we find a saturation value for the field that align particles parallel or antiparallel to the magnetic field, depending on the AMM sign. This value of the magnetic field corresponds to a maximum of spin polarization and to the magnetization independent of B. This is the usual paramagnetic behavior. Beyond this saturation field, the spin polarization, magnetization as well as other thermodynamical quantities, become complex, which points towards a ferromagnetism phase transition.

Since our treatment is based on non-interacting particles, we cannot address the question of whether this limiting value  corresponds or not to a phase transition. To clarify this issue a detailed study of the spin-spin coupling becomes necessary. Let us also remark that the SQM in $\beta$-equilibrium adds new restrictions to the upper bound on the magnetic field. In this case, the polarization of SQM depends on the individual polarization of the constituents and its orientation is related to the AMM sign. The system is so complex that the total polarization is not reached in any direction. Nevertheless, it is possible to find a critical value of the magnetic field beyond which the polarization becomes complex, lacking physical sense.

Our results improve earlier works in three aspects. Firstly, we take into account Pauli paramagnetism in its relativistic version, because the one-particle energy is the solution of the Dirac equation including the AMM. This gives an important contribution to the physics of the system. The upper bound on the magnetic field for each particle is lower than the one obtained classically. Secondly, we consider the anisotropies of the pressures within the MIT bag model. This leads to changes in the behavior of the total energy of the system for strong magnetic fields. Finally, all quarks are assumed to interact with the magnetic field. The most important astrophysical implication of our study is the existence of a limiting value for the magnetic field. For SQM with electrons we find an upper bound on the magnetic field around $8.6 \times 10^{17}$~G. This allows us to conclude that there would not be quark stars with magnetic fields greater than this value.

Another effect associated with the inclusion of the AMM in the particle spectrum is that it can avoid the divergence arising through the lowest Landau level ($n=0$) in the calculation of the surface tension and curvature  for dense matter, and which play a crucial role in the quark droplet nucleation process~\cite{Chakrabarty:1996te}. This means that it is still possible that a first-order phase transition to quark matter can occur~\cite{Suh:2001tr}. A detailed study of this effect and its astrophysical consequences is in progress. The AMM might play a role if Bose-Einstein condensation due to the bosonization of fermions takes place. We expect in that case a ferromagnetic behavior, able to maintain the applied magnetic field self-consistently~\cite{Perez Rojas:2005kt}.

The paper is organized as follows. In Sec.~\ref{sec2} we study the spectrum of constituents of the magnetized SQM: electrons and quarks. Sec.~\ref{sec3} is devoted to the analysis of the thermodynamic properties of the magnetized SQM. In Sec.~\ref{sec4} we establish the requirement for the stability of SQM in $\beta$-equilibrium, and study the spin polarization and its implication for the thermodynamical properties of the system. In Sec.~\ref{sec5} we present our numerical results, including a comparison of the behavior of SQM when AMM are taken into account with the case when the latter are neglected. Finally, our conclusions are presented in Sec.~\ref{sec6}.

\section{Spectrum for the constituents of magnetized SQM}
\label{sec2}

The relativistic spectrum of electrons in the presence of a magnetic field with the inclusion of AMM is obtained from the Dirac Pauli equation
\begin{align}
\left [ \gamma_{\mu} \left(\partial_\mu + i\frac{|e|A_{\mu}}{\hbar c}\right)
-\mu \frac{i}{2\hbar c} F_{\mu\nu}\gamma_{\mu}\gamma_{\nu} +
\frac{m_e c}{\hbar}\right]\psi^{e}=0.
\end{align}
where
\begin{align}
\mu =\frac{\alpha}{2\pi}\frac{e\hbar}{2m_ec}\,,
\end{align}
$e$ is the electron charge, $m_{e}$ is the electron mass, $F_{\mu\nu}$ is the electromagnetic tensor and $\alpha$ is the fine structure constant. Considering a constant uniform magnetic field $B$ in the $x_3$-direction, the energy spectrum is given by\footnote{Hereafter we use the units $\hbar=c=1$.}
\begin{align}
E_{e,n}^{\eta}=m_e \sqrt{x_e^2 + \left( \sqrt{
\frac{B}{B_e^{c}}(2n+1-\eta) + 1} - \eta \alpha\frac{B}{B_e^{c}}
\right)^2},\label{energy_e}
\end{align}
where
\begin{align}
B_e^{c}=\frac{m_{e}^2}{|e|}\,, \quad x_e \equiv \frac{p_3}{m_{e}}\,,
\end{align}
$\eta=\pm 1$ are the eigenvalues corresponding to the orientations of the  particle magnetic moment, parallel or antiparallel to the magnetic field.

An analogous equation can be written for quarks, so that the spectra for all the constituents of SQM have the form

\begin{align}
E_{i,n}^{\eta}=m_i \sqrt{x_i^2 + \left( \sqrt{
\frac{B}{B^c_i}(2n+1-\eta) + 1} - \eta y^{i}B
\right)^2},\label{energy}
\end{align}
with
\begin{align}\label{Bc}
    B^{c}_i=\frac{m_{i}^2}{|e_i|}\,,\quad x_i \equiv \frac{p_3}{m_{i}}\,, \quad y_i=\frac{|Q_i|}{m_i}\,,
\end{align}
$i=(e,u,d,s)$, $e_i$ and $m_{i}$ denote the charges and the masses of the particles, respectively. The quantities $Q_{i}$ are the corresponding AMM of the particles,
\begin{align}
Q_e&=0.00116\mu_B\,,\quad Q_u=1.85\mu_N\,, \nonumber\\
Q_d&=-0.97\mu_{N}\,,\quad Q_s=-0.58\mu_{N}\,,
\end{align}
where
\begin{align}
\mu_{B}&=\frac{e}{2m_e} \simeq 5.79 \times 10^{-15}\, {\rm MeV/G}\,,\nonumber\\
\mu_{N}&=\frac{e}{2m_p} \simeq 3.15\times 10^{-18}\, {\rm MeV/G}.
\end{align}
In our calculations we shall take $m_{u}=m_{d}=5$~MeV and $m_{s}=150$~MeV for the light quark masses. The magnitudes of the so-called critical fields $B^{c}_i$ (when particle's cyclotron energy is comparable to its rest mass) are $B^{c}_e= 4.4\times 10^{13}$~G, $B^{c}_u= 6.3\times 10^{16}$~G, $B^{c}_d=1.3\times 10^{16}$~G and $B^{c}_{s}=1.1\times 10^{19}$~G.

It can be seen from the spectra (\ref{energy}) that, besides of the quantization of their orbits in the plane perpendicular to the magnetic field, charged particles with AMM undergo the splitting of the energy levels with the corresponding disappearance of the spectrum degeneracy. For the non-anomalous case, $Q_{i}=0$, the minimum energy is independent of the magnetic field strength and the magnetic field only quantizes the kinetic energy perpendicular to the field. In this situation, the energy is degenerate for Landau levels higher than zero. States with spin parallel or antiparallel to the magnetic field orientation ($\eta=\pm 1$) have the same energy. However, the anomalous case, $Q_i\neq 0$, removes this degeneracy. In the latter case, the rest energy of the particles depends on the  magnetic field strength. The ground state energy is
\begin{align}
E_{i,0} = m_i  \left(1-y_i B\right) .\label{energy1}
\end{align}

The above equation leads to the appearance of a threshold value for the magnetic field at which the effective mass
vanishes, $m_i \sim |Q_i| B$. The thresholds of the field, $B^s_i=1/y_i$,  for all the constituents of the SQM are
given by
\begin{align}
B^{s}_e&=7.6\times 10^{16}\,\text{G},\quad B^{s}_u=8.6\times 10^{17}\,\text{G},\nonumber\\
B^{s}_d&=1.6\times 10^{18}\,\text{G}, \quad B^{s}_s=8.2\times 10^{19}
\,\text{G},
\end{align}
that are smaller than the ones obtained when the classical AMM contribution is considered~\cite{Chakrabarty:1996te}.

The meaning of the ground state energy value for QED was discussed long  time ago~\cite{Chiu:1968}. Nevertheless, it was not emphasized enough the fact that, due to the degeneracy of the orbit center, such ground state level may be populated by a larger number of particles. The expression (\ref{energy1}) suggests that the energy of the particles becomes smaller than the corresponding one for the antiparticles, with the consequent creation of pairs. The sign of the energy state is an invariant property for particles and antiparticles. This also means that positive and negative energy levels of electrons will never cross each other, i.e. it exists a non-crossing property. The spontaneous pair creation in a magnetic field is forbidden. Thus, for individual particles, the correct meaning of this ``critical" field is that it corresponds to an upper bound.

Let us remark that in the SQM scenario all the constituents interact with the magnetic field and are obliged to satisfy the equilibrium conditions. Under such constraints, it turns out that the dominant threshold field comes from $u$ quarks, thus leading to the upper bound $B \lesssim 8.6 \times 10^{17}$~G (see section \ref{sec5} below). This result has an important astrophysical consequence, since the bound for SQM can be also extrapolated to the SQS scenario. If  SQS exist, the maximum magnetic field strength that they could support  would be around the above bound, i.e. $10^{18}$~G.

\section{Thermodynamic properties of magnetized SQM with AMM}
\label{sec3}

The MIT bag model is appropriate for the study of magnetized quark matter~\cite{Perez Martinez:2005av}. In that model, confinement is guaranteed by the bag and quarks are considered as a Fermi gas of noninteracting particles. Under these assumptions, it is possible to study the thermodynamical properties of a quark gas in a strong magnetic field. In this section we investigate  the thermodynamical properties of the SQM when the AMM is included.

The inclusion of AMM and the consequent loss of degeneracy implies that the sum over Landau levels is replaced by two sums
\begin{align}
\sum^{n^i_{max}}_{n=0}(2-\delta_{0n}) \Rightarrow
\sum^{n^i_{max}}_{n=0}\sum_{\eta}\,.
\end{align}
For each thermodynamical quantity, the summation over the  spin orientation  leads to two contributions, corresponding to particles with the spin aligned parallel or antiparallel to the magnetic field. Moreover, since particles have positive or negative AMM, they have different preferences in the  spin orientation with respect to the magnetic field. As we shall see below, this has important consequences in the EOS of the system. The most relevant comes from the lowest energy ground state, which depends on the strength of the magnetic field and it could be zero (cf. Eq.~(\ref{energy1})).

For a degenerate magnetized SQM, where only particles contribute to the thermodynamical potential and temperature can be formally taken as zero, the expression for the thermodynamical potential  can be written in the form~\cite{Perez Martinez:2005av}
\begin{align} \label{potential}
\Omega_i&={\mathcal M}^0_{i}B\sum_{n}\left(\Omega^{+}_ i+\Omega^{-}_i\right),\quad {\mathcal M}^0_{i} = \frac{d_{i}e_i m_{i}^2}{4\pi ^2}\,,\nonumber\\
\Omega^{\pm}_i&=-x_{i}g^{\pm}_{i}
+ h^{\pm\,\,2}_{i} \ln\frac{x_{i} + g_{i}^{\pm}}{h^{\pm}_{i}}\,,
\end{align}
where $i=e,u,d,s$ and $d_i$ is a degeneration parameter ($d_e=2, d_{u,d,s}=6$). We have defined $x_{i}=\mu _{i}/m_{i}$, where $\mu_i$ is the chemical potential; $h_i$ and $g_i$ are dimensionless functions given by\footnote{To simplify the notation, from now on we omit the Landau level subscript $n$ in all quantities.}
\begin{align}
g_{i}^{\eta} = \sqrt{x_{i}^2-h_{i}^{\eta}\;^2}\,,
\end{align}
\begin{align}
h_{i}^{\eta} =
\sqrt{\frac{B}{B^{c}_i} (2n + 1-\eta) +1} -\eta y_{i}B\,. \label{hE}
\end{align}
The sum over the Landau levels $n$ is up to $n_{max}^{i}$ given by the expression
\begin{align}
n_{max}^i = I\left[\frac{(x_{i} +  \eta y_iB)^2 -
1}{2B/B^{c}_i}\right],\label{Landau}
\end{align}
where $I[z]$ denotes the integer part of $z$.

The density of particles, defined as $N=\sum_i N_i$ with
$N_i=\frac{\partial\Omega_{i}}{\partial\mu_i}$ gives
\begin{align}\label{density}
 N_i= N^0_{i}\frac{B}{B^c_{i}}\sum_{n} \left( g_{i}^{+}
+g_{i}^{-}\right),\quad
N_{i}^0 =  \frac{d_i m_{i}^3}{2\pi^2}.
\end{align}
The magnetization of the SQM is ${\mathcal M}= \sum_i{\mathcal
 M}_i,$ with ${\mathcal M_i}=-\frac{\partial\Omega_i}{\partial B}$. We find
\begin{align} \label{magnet}
{\mathcal M}_i&= {\mathcal M}^0_{i}\sum_{n}\left({\mathcal M}_{i}^{+}+{\mathcal M}_{i}^{-}\right),\nonumber\\
 {\mathcal M}_{i}^{\pm}&= g_{i}^{\pm}  x_{i} -
\left(h_{i}^{\pm\,2} + 2h_{i}^{\pm} \gamma_i ^{\pm}\right)
\ln\frac{x_{i} + g_{i}^{\pm}}{h_{i}^{\pm}},
\end{align}
with
\begin{align}
\gamma^{\eta}_i=\frac{(2n+1-\eta)B}{2B^c_i\sqrt{B/B^c_i(2n+1-\eta)+1}}-\eta
y_i B\,.
\end{align}

We may also calculate the magnetic susceptibility $\chi=\sum_i \chi_i$, which is defined as $\frac{\partial{\mathcal M}_i}{\partial B}$ and can measure if a phase transition takes place or not. The resulting expression is
\begin{align}
\chi_i= \frac{{\mathcal M}^0_{i}}{B} \sum_{n}\left( \chi^{+}_i
+\chi^{-}_i\right),
\end{align}
where
\begin{align}
\chi_i ^{\pm} &=\frac{2\gamma_i^{\pm\,\,2}
x_i}{g^{\pm}_{i}}-\left(4\gamma_i^{\pm} h_{i}^{\pm}+
2\gamma_i^{\pm\,\,2} \right)\ln\frac{x_i + g_{i}^{\pm}}{h_{i}^{\pm}}
\nonumber \\
&\,\quad\quad + \frac{h_{i}^{\pm}(2n+1 \mp 1)^2
(B/B^{c}_i)^2}{2[(2n+1 \mp 1)B/B^c_{i}+1]^{3/2}} \ln\frac{x_i +
g_{i}^{\pm}}{h_{i}^{\pm}}\,.
\end{align}

We notice that the magnetization and magnetic susceptibility are not linear functions of $B$. Moreover, the requirement for the ground state energy $ E_{i,0} \geq 0$ is equivalent to the condition $h^{\eta}_{i}(n=0) \geq 0$ (see Eq.~(\ref{hE})). Since all the thermodynamical quantities depend on $h^{\eta}_{i}$, they loose their  physical meaning when $h^{\eta}_{i} < 0$. Therefore, the condition $h^{\eta}_{i}\geq 0$ means that the system cannot admit a value of the magnetic field greater than $B_i^s$, which indicates that some phase transition occurs. The magnetization reaches a value independent of the magnetic field, as it occurs for a paramagnetic system, but beyond this limit it becomes complex and could be associated to a ferromagnetic transition.

\subsection{Pressure and energy density}

Let us now write down the expression for the anisotropy of pressures and for the energy density of SQM when the AMM is included. The energy density, $U$, for the gas of $i$-particles can be obtained from the energy-momentum tensor~\cite{Perez Martinez:2005av}, leading to
\begin{align}
U=\sum_i U_i\,,  \quad U_i=\Omega_i +x_iN_i\,.
\end{align}
Evaluating this expression we find
\begin{align}
 U_i&={\mathcal M}^0_{i}B\sum_{n}\left(U^{+}_i+U^{-}_i\right)\,, \nonumber\\
U^{\pm}_i&=x_qg^{\pm}_i+h_{i}^{\pm\,\,2}\ln\frac{x_i+g_i^{\pm}}{h_{i}^{\pm}}\,.
\end{align}
The pressures are obtained from the expressions
\begin{align}
P_{\perp} &= \sum_i P_{i\,\perp}\,, \quad
P_{i\,\perp}=-\Omega_i-{\mathcal M}_iB\,,\nonumber\\
P_{\parallel}&=\sum_i P_{i\,\parallel}\,,\quad
P_{i\,\parallel}=-\Omega_i\,.
\end{align}
Using Eqs.~(\ref{potential}) and (\ref{magnet}) we find
\begin{align}
P_{i\,\parallel}&={\mathcal M}^0_{i}B \sum_{n} \left(P_{i\,\parallel}^{+} + P_{i\,\parallel}^{-}\right)\,,\nonumber\\
 P_{i\,\parallel}^{\pm} &=  x_{i}g_{i}^{\pm} - h^{\pm\,\,2}_{i}\ln\frac{x_{i} +
g^{\pm}_{i}}{h^{\pm}_{i}}\,,
\end{align}
and
\begin{align}
P_{i\,\perp} &= {\mathcal M}^0_{i}B \sum_{n}\left( P_{i\,\perp}^{+} +
P_{i\,\perp}^{-}\right)\,,\nonumber\\
P_{i\,\perp}^{\pm} &= 2h_{i}^{\pm} \gamma_i^{\pm}
\ln\frac{x_{i} + g_{i}^{\pm}}{h_{i}^{\pm}}.
\end{align}

\section{The stability condition for magnetized SQM}
\label{sec4}

In this section we study the stability condition of the SQM in a strong magnetic field. In the context of the MIT bag model and in the absence of a magnetic field, the stability condition for SQM means to study the equation
\begin{align}
P_T+ B_{bag}= \sum_iP_i\label{P_bag}\,,
\end{align}
together with the total energy
\begin{align}
U_T-B_{bag}=\sum_i U_i\label{U_bag}\,,
\end{align}
under the condition $P_T=0$.

In the presence of a strong magnetic field, the bag pressure $B_{bag}$ has an anisotropic form that depends on
the $B$-direction in space,
\begin{align}
B_{bag}^{\parallel} &\equiv \sum_iP_{i\,\,\parallel}\,,\nonumber\\
B_{bag}^{\perp} &\equiv \sum_i P_{i\,\,\perp}\,.
\end{align}
Since the magnetization is always a positive function, this anisotropy in the pressures implies
$P_{\perp}<P_{\parallel}$. Thus, the stability  condition for strong fields changes from $P_T=0$ to $P_T^{\perp}=0$ or, equivalently,
\begin{align}
\sum_i\Omega_i =-\sum_i {\mathcal M}_i B\,.
\end{align}
The total energy becomes
\begin{align}
U_T=\sum_i (-{\mathcal M}_i B +x_i N_i)\,.
\end{align}
For weak fields $P_{\perp}=P_{\parallel}\,$ and we get the expression
\begin{align}
U_T=\sum_i x_i N_i\,,
\end{align}
which is in agreement with~\cite{Ghosh:1999hh} where anisotropies due to strong magnetic fields have not been considered.

\subsection{Paramagnetism response: spin polarization}

In Ref.~\cite{Chakrabarty:1996te} the Landau diamagnetism related to charged particles in a magnetic field was studied,  treating classically the relativistic behavior and Pauli paramagnetism associated with the inclusion of AMM. Our starting viewpoint is different. We consider a relativistic equation of motion, taking into account both contributions from a relativistic point of view. In this sense, our treatment is more robust since the  AMM is included in the spectrum of particles as a relativistic effect.

As we have already shown, the thermodynamical quantities for each constituent of SQM have two terms, related to particles with spin up and down orientations.  The number density is not excluded from this fact,  so it is important to study its behavior because it gives us information about the spin polarization of the system. Let us rewrite Eq.~(\ref{density}) in the form
\begin{align}
    N_i=N_i^{\uparrow} +N_i^{\downarrow}\,.
\end{align}
In the absence of a magnetic field or when $B \rightarrow 0$ we can see that $N_i^{\uparrow} =N_i^{\downarrow}$. On the other hand, in the presence of a magnetic field, the relation for the number density implies  the existence of a magnetic field strength threshold for which complete saturation of each constituent of the SQM occurs. Whether or not a complete saturation for all the particles involved in the system is attained will depend on the SQM equilibrium conditions.

We can define the spin polarization rate as
\begin{align}
S_{p}^{i\uparrow\downarrow} =\frac{N_{i}^{\uparrow\downarrow
}}{N_{i}^{\uparrow}+ N_{i}^{\downarrow}}.\label{polarization}
\end{align}
When $B_i^s$ is reached,  we have the following condition for each constituent particle,
\begin{align}
    S^{i\uparrow}_p =1\,,\quad S^{i\downarrow }_p=0\,,
\end{align}
which means that $N^{\uparrow}_i$ is maximum and $N^{\downarrow}_i=0$. The threshold field value $B_{i}^{s}$ saturates the system and aligns all particles parallel or antiparallel to the magnetic field. This alignment depends on the sign of the AMM of each particle. From the thermodynamical point of view, this behavior could be understood as a paramagnetic response of the system and it makes an important difference. Beyond this magnetic field value all the thermodynamical quantities become complex for a pure gas of particles.

Let us recall that the conditions of $\beta$-equilibrium and charge neutrality  add new restrictions to the threshold values $B_i^s$. These values were computed numerically for all the chemical potentials of the SQM constituents to study, in the next section, the spin polarization in the regime of a strong magnetic field. As it turns out, the $\beta$-equilibrium and charge neutrality are satisfied only for magnetic field values below certain threshold. Above this, the number density of electrons and $u$ quarks are fixed, independently of the magnetic field, and $\beta$-equilibrium will require $\mu_e <0$. Our numerical results confirm that the main contribution to the threshold field comes from $u$ quarks,  which imply the upper bound $B \lesssim B_u^s = 8.6 \times 10^{17}$~G.

\section{SQM in $\beta$-equilibrium and charge neutrality: numerical results}
\label{sec5}

In this section we perform a complete numerical study with the aim to determine all the relevant thermodynamical quantities for SQM and discuss its stability, taking into account $\beta$-equilibrium and charge neutrality. This requires the solution of a system of equations to obtain the chemical potentials of all the species involved in the system. If SQM exists in the core of neutron stars or forms itself a SQS, weak processes will be responsible for the appearance of the $s$ quarks. Once this occurs, the equilibrium among the constituents will be dynamically established.

The three ingredients for the SQM in equilibrium are $\beta$-equilibrium, charge neutrality and the conservation of the baryonic density $n_B$:
\begin{align}\label{SQMeqs}
&\mu_u+\mu_e=\mu_d\,,\quad \mu_d=\mu_s\,,\nonumber\\
&2N_u-N_d-N_s-3N_e=0\,,\nonumber\\
&\frac{1}{3} (N_u+N_d+N_s)=n_{B}\,.
\end{align}
Here we assume that there is no neutrino trapping in the system, so that they do not play any role on the $\beta$-equilibrium conditions. For a given baryon density  (we take $n_B = 2.5\, n_0=0.4$~fm$^{-3}$) and magnetic field strength, these equations together with Eqs.~(\ref{density}) allow us to determine the chemical potentials and evaluate all the thermodynamic properties of the system. At the end of this section we shall comment on the dependence of our results with the variation of $n_B$.

\begin{figure}[t]
\includegraphics[width=8.5cm]{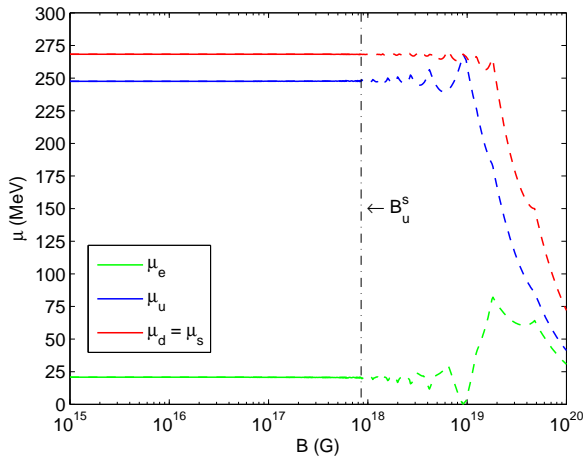}
\caption{(Color online) Chemical potentials $\mu_i$ for SQM as functions of the magnetic field strength $B$ with
 (solid lines) and without (dashed lines) AMM. The vertical dot-dashed line corresponds to the threshold
 value $B_u^s \simeq 8.6\times 10^{17}$~G.}\label{Fig1}
\end{figure}

\begin{figure}[t]
\includegraphics[width=8.5cm]{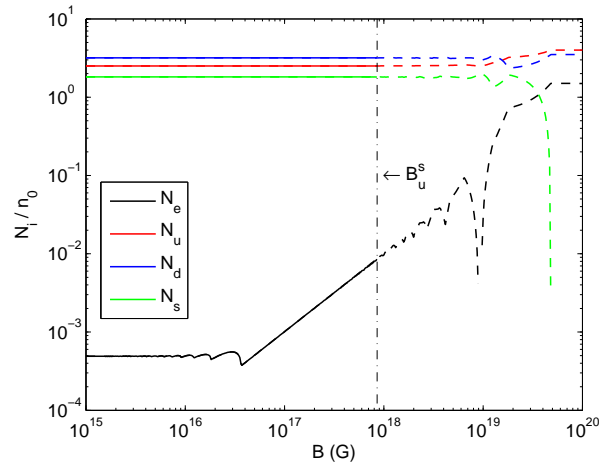}
\caption{(Color online) The variation of the number densities with the magnetic field. The solid
(dashed) lines correspond to the case with (without) AMM. The baryon density is fixed at the value
$n_B=0.4$~fm$^{-3}$.}\label{Fig2}
\end{figure}

\begin{figure}[t]
\includegraphics[width=8.5cm]{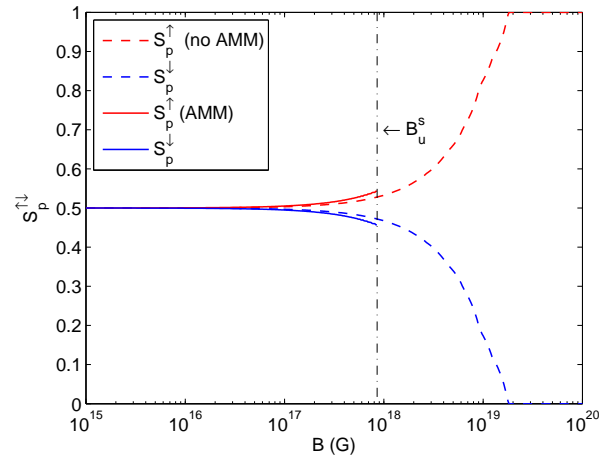}
\caption{(Color online) Spin polarization for SQM as a function of the magnetic field strength $B$ with
(solid lines)  and without (dashed lines) AMM.}\label{Fig3}
\end{figure}

In Fig.~\ref{Fig1} we show the chemical potentials $\mu_i$, i.e. the solution of Eqs.~(\ref{SQMeqs}), as functions of the magnetic field $B$.  Note that the chemical potentials remain practically constant up to the threshold $B_u^s$ around $8.6\times 10^{17}$~G, which corresponds to the upper bound on the magnetic field determined by the $\beta$-equilibrium  condition. The variation of the number densities for all the SQM constituents  with the magnetic field $B$ is shown in Fig.~\ref{Fig2}. Comparing this variation with and without the AMM inclusion, we can see that for relatively small values of the magnetic field, $B \lesssim 10^{16}$~G, all the number densities remain practically constant. At around the magnetic field value of $4\times 10^{16}$~G, the electron density $N_e$ start to increase with $B$, whereas the quark densities remain almost constant up to field strengths of $10^{19}$~G. Above this value the $s$-quark density $N_s$ decreases with $B$ and becomes negligibly small. The oscillations due to the presence of Landau levels can be seen for the case when the AMM is not considered. Clearly, the AMM curves are bounded by the upper bound $B_u^s$; for fields greater than this value all the number densities become complex.

\begin{figure}[t]
\includegraphics[width=8.5cm]{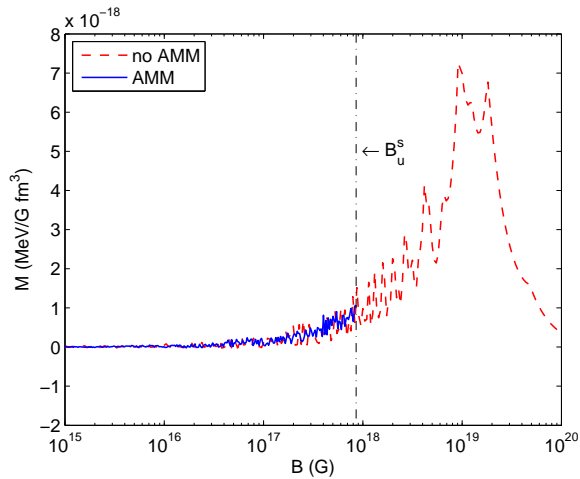}
\caption{(Color online) Magnetization as a function of the magnetic field with (solid line) and without
(dashed line) AMM.}\label{Fig4}
\end{figure}

\begin{figure}[t]
\includegraphics[width=8.5cm]{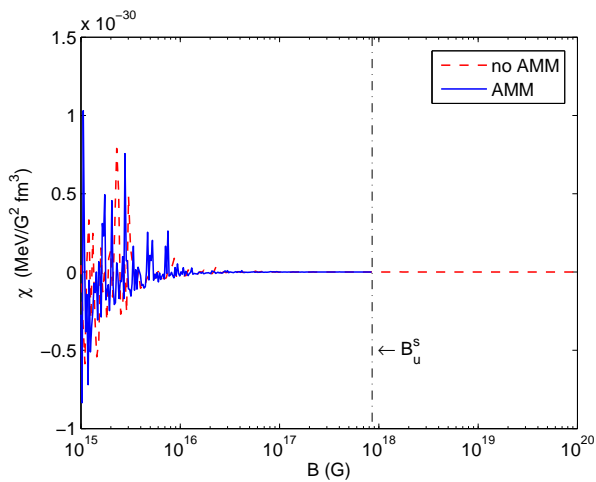}
\caption{(Color online) Susceptibility as a function of magnetic field for the two cases considered:
without the inclusion of AMM (dashed line) and with their inclusion (solid line).}\label{Fig5}
\end{figure}

\begin{figure}[t]
\includegraphics[width=8.5cm]{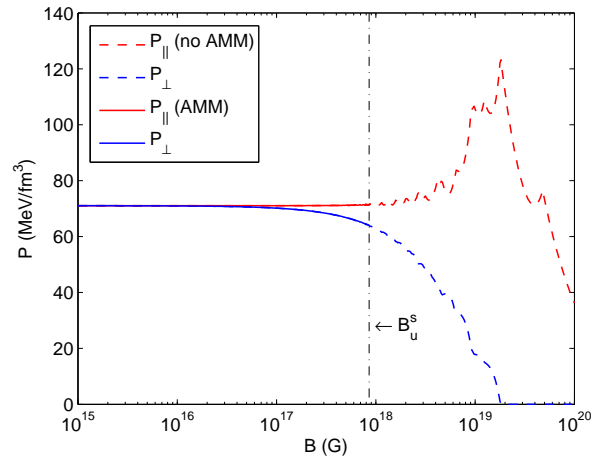}
\caption{(Color online) Anisotropy of the SQM pressures as functions of the magnetic field strength.
The cases with (solid lines) and without (dashed lines) AMM have been considered.}\label{Fig6}
\end{figure}

\begin{figure}[t]
\includegraphics[width=8.5cm]{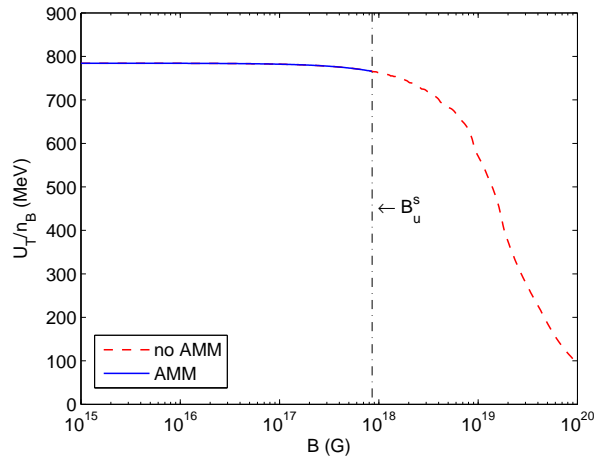}
\caption{(Color online) Energy per baryon versus $B$. As in previous figures,
the dashed line correspond to the case when the AMM is not included, whereas
the solid line takes into account the AMM of all the SQM constituents.}\label{Fig7}
\end{figure}

The total spin polarization of SQM, $S_{p}^{\uparrow\downarrow}= \sum_i S_{p}^{i\uparrow\downarrow}$, is plotted in Fig.~\ref{Fig3} as a function of the magnetic field. We see that the spin polarization of the system increases with the increasing of the magnetic field. When the AMM are not included, a total polarization is achieved for $B \simeq 2 \times 10^{19}$~G. On the other hand, when the AMM are taken into account the system cannot reach a total spin polarization since for values greater than $B_u^s$ the density number becomes complex.

The behavior of the magnetization ${\mathcal M}$ is depicted in Fig.~\ref{Fig4}. It is always a positive quantity for fields greater than $10^{16}~G$. It also exhibits the so-called de Haas-van Alphen oscillations, with increasing amplitude as $B$ increases. This is even more noticeable when the AMM is not considered as higher values of $B$ are allowed. The corresponding magnetic susceptibility $\chi$ is presented in Fig.~\ref{Fig5}. It shows the paramagnetism behavior of SQM for fields larger than $10^{16}~G$. Below this magnetic field strength, $\chi$ has an oscillating behavior. The upper bound on the magnetic field around $8.6\times 10^{17}\,G$ encloses a phase transition of second type, because at this value all particles align parallel or antiparallel to the magnetic field with a positive value of the magnetization.

\begin{figure}[h]
\includegraphics[width=8.5cm]{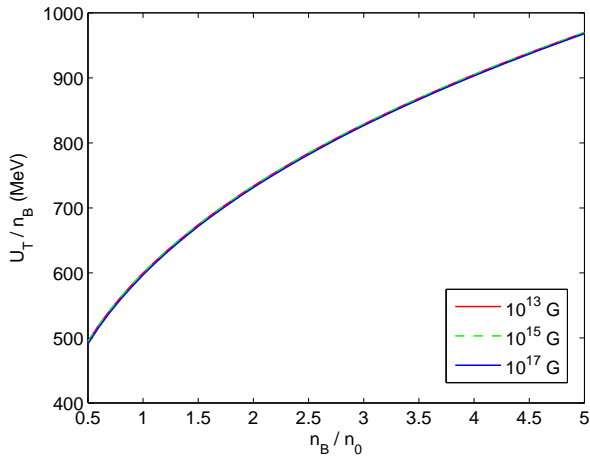}
\caption{(Color online) Energy per baryon versus $n_B$ with AMM included
for $B=10^{13},\,10^{15},\,10^{17}$~G.}\label{Fig8}
\end{figure}

\begin{figure}[t]
\includegraphics[width=8.5cm]{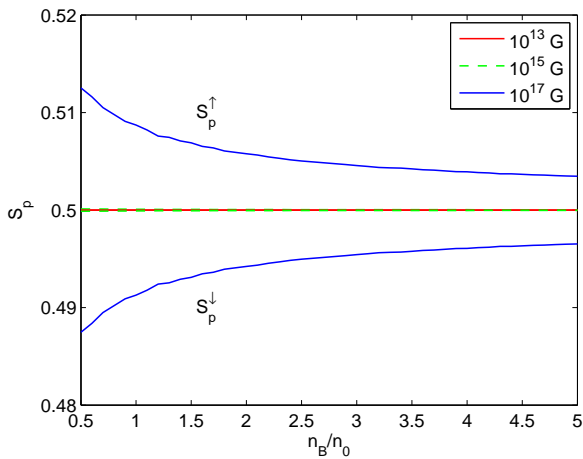}
\caption{(Color online) Total spin polarization versus $n_B$ with AMM
included for $B=10^{13},\,10^{15},\,10^{17}$~G.}\label{Fig9}
\end{figure}

Let us now consider the pressure and total energy of the system. We plot in Fig.~\ref{Fig6} the behavior of the pressure with the variation of the magnetic field. The system persists being anisotropic when the AMM are considered. For vanishing AMM, the perpendicular component of the pressure $P_{\perp}$ goes to zero at about $2 \times 10^{19}$~G, when the total spin polarization is reached. On the other hand, the inclusion of the AMM forbids fields above the threshold $B_u^s$. Thus, for SQM the anisotropy in the pressures is relatively small, i.e, $P_{\perp} \simeq P_{\parallel}$.

Fig.~\ref{Fig7} shows the behavior of the total energy per baryon with the magnetic field. We have plotted two curves: particles with AMM and without it. The figure confirms that SQM is stable up to the corresponding field threshold. For $B \lesssim B_u^s$ the energy per baryon remains almost constant, and decreases for higher values of the magnetic field.

To conclude this section, let us comment on the variation of the thermodynamic properties with the baryon density $n_B$. In Fig.~\ref{Fig8} we show the dependence of the total energy per baryon with the baryon density for three different values of the magnetic field, $10^{13}$, $10^{15}$ and $10^{17}$~G, taking into account the AMM. As the baryon density increases, the total energy of the system increases too. It also turns out that SQM is energetically more stable in the presence of a strong magnetic field\footnote{In the absence of a magnetic field, and for a given value of $n_B$, the SQM energy per baryon is always higher than the value obtained when $B\neq0$~\cite{Chakrabarty:1996te}.}. For a fixed value of $n_B$, we remark that  there are no significant changes as the magnetic field varies (the three curves are almost indistinguishable, as can be seen from the figure). The same is observed for other thermodynamical quantities. Nevertheless, the total spin polarization increases with the increasing of the magnetic field strength and the decreasing of the baryon density $n_B$, as can be seen in Fig.~\ref{Fig9}.

\section{Conclusions}
\label{sec6}

We have studied the magnetized SQM in $\beta$-equilibrium in the presence of a strong magnetic field. We have taken into account the Landau diamagnetism related to the quantization of the Landau levels as well as the Pauli paramagnetism, due to the presence of AMM for all the constituents of SQM. The influence of the paramagnetism in the system is more relevant than the diamagnetism because it is responsible for the upper bound on the magnetic field found for the SQM system. This bound is lower than the one obtained classically~\cite{Chakrabarty:1996te}. Furthermore, it implies that a phase transition should occur at this value, because all individual particles are aligned parallel or antiparallel to the magnetic field (depending on the AMM sign) in the ground state of the energy.

For SQM in $\beta$-equilibrium and with neutral charge the situation is mathematically complex. The condition of $\beta$-equilibrium implies an upper bound on the magnetic field, $B \lesssim B_u^s = 8.6 \times 10^{17}$~G. Above this value, the chemical potential of electrons becomes negative and all the thermodynamical quantities loose their  physical meaning. As a consequence, a total spin polarization is not achieved, in contrast with the case without AMM, where such a polarization is reached for fields $\sim 10^{19}$~G.

From the quantum statistical point of view the lowest energy states with AMM contain important physical consequences: for particles with mass $m_i$ and anomalous magnetic moment $Q_i$, the magnetic field has a critical value given by the expression $B_i^s \sim m_i/|Q_i|$. It remains to clarify if, for a magnetic field strength of this order, quark matter undergoes a phase transition. This question deserves particular study. On the other hand, it becomes clear that the stability condition of SQM is modified in the presence of a strong magnetic field.

In this work we have shown the differences that the AMM introduces in all thermodynamic properties. We have concluded that magnetized SQM with AMM is stable as it is in the case when no AMM is introduced. In both cases, magnetized SQM is more stable than SQM without a magnetic field. The pressures preserve the anisotropies found in~\cite{Martinez:2003dz} for pure neutron and electron gases in strong magnetic fields. Nevertheless, a significant anisotropy cannot be reached due to the presence of the AMM.

In summary, we have obtained a threshold value for the magnetic field which is equal to the saturation field for the $u$ quarks. This value of the magnetic field is due to the restrictions of $\beta$-equilibrium and it limits all the thermodynamical quantities. From the astrophysical point of view, our conclusions imply that, if SQS exist, they cannot support magnetic fields greater than $10^{18}$~G.

As mentioned before, if there is bosonization, which is otherwise expected, for instance, in the form of di-quarks,  the model of Bose condensation developed in~\cite{Perez Rojas:2005kt,Perez Rojas:1995ja} could  be applied. For SQM, in that case, the ferromagnetic phase transition  due to AMM would be guaranteed for fields of the order $\sim 10^{18}$~G. This could indicate another type of phase transition for SQM, in addition to the CFL~\cite{Alford:1998mk} (or mCFL~\cite{Ferrer:2005vd}) phases.

\section*{Acknowledgments}

We are grateful to H. Mosquera Cuesta for reading the manuscript and sending us suggestions. A.P.M. thanks D. Mart\'{\i}nez for discussions, and the ICTP and CFTP-IST~(Lisbon, Portugal) for their hospitality. M.O. thanks CLAF for the hospitality. The work of R.G.F. has been partially supported by \emph{Funda\c{c}\~{a}o para a Ci\^{e}ncia e a Tecnologia} (FCT, Portugal) through the project PDCT/FP/63912/2005. The work of A.P.M. and H.P.R. has been supported by \emph{Ministerio de Ciencia, Tecnolog\'{\i}a y Medio Ambiente} under the grant CB0407. A.P.M and H.P.R. acknowledge the support of the ICTP Office of External Activities.

\end{document}